\documentstyle[12pt]{article}
\def\eg{{\frenchspacing e.{\thinspace}g. }}
\def\cf{{\frenchspacing\it cf. }}
\def\etal{{\frenchspacing\it et al.}}
\def\simlt{\hbox{ \rlap{\raise 0.425ex\hbox{$<$}}\lower 0.65ex\hbox{$\sim$} }}
\def\simgt{\hbox{ \rlap{\raise 0.425ex\hbox{$>$}}\lower 0.65ex\hbox{$\sim$} }}
\def\pasp{\rm PASP\/}
\def\apj{\rm ApJ\/}
\def\pasj{\rm PASJ\/}
\def\apjs{\rm ApJS\/}
\begin{document}

\title{Blue \ Stragglers \ as \ Tracers \ of \ Globular \ Cluster \ Evolution}
\author{Piet Hut \\
Institute for Advanced Study, Princeton, NJ 08540, U.S.A.}
\maketitle

\begin{abstract}
Blue stragglers are natural phenomena in star clusters.  They
originate through mass transfer in isolated binaries, as well as
through encounters between two or more stars, in a complex interplay
between stellar dynamics and stellar evolution.  While this interplay
cannot be modeled quantitatively at present, we will be able to do so
in one or two years time.  With this prospect, the present paper is
written largely as a preview.

Star clusters with a high central density contain an ecological
network of evolving binaries, affected by interactions with passing
stars, while in turn affecting the energy budget of the cluster as a
whole by giving off binding energy.  The energy liberated can balance
the losses from the central regions by escaping stars as well as
the energy lost by a heat flow toward the relatively colder cluster halo.

The `gravitational fusion' of single stars into binaries, triples, and
collisional merger products proceeds via a complex reaction network.
Although we are beginning to chart the activity in some of the major
channels, there are still major uncertainties, and consequently our present
knowledge of blue straggler formation and evolution is largely qualitative.

Quantitative progress is contingent on advances in the following four
areas: 1) observations of fundamental cluster parameters, such as the
mass spectrum and the abundance of blue stragglers in the central regions;
2) availability of faster computers to model clusters on a star-by-star basis,
with a Teraflops speed being desirable; 3) further improvement in N-body codes
to make full use of such speeds; and 4) the development of consistent
`evolutionary recipes' to treat the interplay between stellar
evolution and stellar dynamics.  These requirements can be met in the
next couple years.
\end{abstract}

\section{Introduction}

Halfway between the study of individual stars (by now a relatively
well--under-stood area) and that of galactic nuclei (still not very well
understood) lies the study of star clusters.  If we could not reach a
detailed understanding of the basic structure and evolution of
galactic and globular clusters, a quantitative modeling of active
galactic nuclei would be even more remote.

With this motivation, it is rewarding to focus our attention on the
densest and richest clusters available near our galaxy, the globular
clusters, and especially their central areas where close encounters
and even physical collisions between stars are not infrequent.  Some
of these encounters can produce exotic objects such as X-ray binaries
and millisecond pulsars, but most encounters will involve
garden-variety main-sequence stars.  Judging from the numbers of
exotic objects, as well as from back-of-the-envelope estimates,
globular clusters must have formed a stage for thousands of stellar
collisions, as first shown by Hills \& Day (1976).  What did these
collisions produce?

When two main-sequence stars collide in a low-velocity-dispersion
environment provided by a globular cluster, very little mass can
escape, since the specific kinetic energy at infinity is typically
three orders of magnitude smaller than the escape energy at the
surface of a star.  Therefore, the merger remnant will consist of a
simple addition of the masses of the two stars, albeit in a rather
excited stage at first.  After the initial oscillations have damped
out on a dynamical time scale, and the thermal excess has been
radiated away on a thermal time scale, the resultant star is expected
to resume a rather normal appearance.

Depending on the details of the collision and the prior evolutionary
state of the stars, the merger product may have an excess rotation and
unusual abundances (Bailyn 1992), but by and large we will find
ourselves simply with an overweight star, possibly of a type we don't
expect to encounter any more at the present evolutionary state of the
cluster -- in this case we will have produced a blue straggler.

\bigskip
\noindent
{\it Star-by-Star Modeling of Star Clusters}
\medskip

To summarize: if want to probe the collisional history of dense
stellar systems, and if we want to obtain optimal statistics, we
should look for the products of collisions between ordinary stars.
They come in two varieties.  Those merger remnants that are less
massive than the main sequence turn-off are buried in the HR diagram
like needles in a hay stack.  But those that are more massive do stand
out as `blue stragglers' -- that is, if we can resolve the region of
interest down to the level of the main sequence.  Now that this is
possible, with the Hubble Space Telescope in even the densest cluster
cores, it is time to roll up our sleeves and get serious with our
modeling efforts.

Not that we have not been serious so far.  The evolution of star
clusters has approached a state of maturity comparable to that of
stellar evolution three decades ago.  We have begun to understand the
physics driving core collapse and post-collapse evolution, and we are
on the brink of building detailed models necessary for comparisons
with observations.  Since we have just published an extensive review
of recent modeling efforts (Hut \etal\ 1992, section 3), we can limit
ourselves to a brief summary-style review in \S\S2,3.  From \S4
onwards, the paper is presented as a preview.

Another reason to look at the future, rather than the past, is that
our present modeling efforts are simply not yet capable to meet the
challenges posed by the state-of-the art observations.  As discussed
by Hut \etal, Fokker-Planck simulations cannot handle binaries
adequately, while $N$-body methods still lack the necessary
computational speed.  In fact, at present three barriers still
separate us from the goal of reaching parity with observational
advances, as discussed below in \S\S 4-6.  What is most exciting, and
what forms the central theme of this preview, is that we now have a
firm time table for scaling these barriers, namely during the next one
or two years.

\bigskip
\noindent
{\it Overview}
\medskip

Rolling up our sleeves is literally an appropriate expression for
theorists who want to follow the evolution of globular clusters.
Since we need a Teraflops-month to do so, we can either wait till the
next millennium when that type of compute has become affordable, or we
can build our own star-cluster machine.  Fortunately, some
enthusiastic astronomers in Tokyo have started to do just that, and are
expected to produce the necessary cycles in one or two years, as will
be discussed in \S 4.

However, speed alone won't do, and we also need to extend our present
software, to be able to handle the extreme problems of disparate
length scales and time scales (by relative factors of up to $10^{20}$).
Current efforts in that directions are described in \S 5.

Handling $10^5$ point masses, although a good start towards globular
cluster modeling, by itself will not enable meaningful comparisons
with observations.  We really need to take into account the intricate
interplay between stellar evolution and stellar dynamics.  A review of
the general problem is given in \S 3, and a preview of our current
modeling efforts is presented in \S 6.

With proper speed, integration algorithms, and stellar evolution
recipes all in hand, a year or so from now, we will have to sort out
the processes of interests from among the terabytes of data generated
in star-by-star cluster simulations.  This is the topic of \S 7.  \S 8
sums up.

Before reviewing and previewing the various cluster evolution modeling
efforts, we first summarize in \S 2 the physical principles underlying
our understanding of the dynamics of dense star clusters, and in \S 3
their application to globular clusters.

\section{Gravitational Fusion}

During the last ten years, enormous progress has been achieved in
globular cluster dynamics.  We now understand the phenomena of core
collapse and post-collapse gravothermal oscillations, as well as the
important role that primordial binaries play.  Since there are various
recent reviews that cover these topics (\eg Goodman 1992, Hut 1992,
Hut \etal\ 1992), I will simply outline the basic physical principles of
cluster stellar dynamics here, before discussing the connections
with stellar evolution.

Double stars play a central role in cluster dynamics.  If their
orbital speed exceeds that of the velocity dispersion of the single
stars, the tendency toward energy equipartition during encounters will
transfer some of the internal kinetic energy to passing stars.
Doing so, energy conservation causes them to shrink, while the
negative heat capacity of self-gravitating systems causes them to heat
up further, to higher orbital speeds (Lynden-Bell's `donkey effect':
trying to slow down particles in a Kepler orbit speeds them up, and
{\it vice versa}).

The `gravitational fusion' of single stars into double stars is thus
one mechanism that can heat a cluster, in order to balance the energy
losses due to the evaporation of stars and the heat flow through the
cluster toward the colder halo.  Some of the analogies with nuclear
fusion in stars, as well as a derivation of binary distribution
functions as the classical limit of the hydrogen atom, are reviewed by
Hut (1985).  Some of the most detailed studies of gravitational
scattering, the mechanism of gravitational fusion, can be found in
Heggie \& Hut (1993) and Goodman \& Hut (1993) and references therein.

\bigskip
\noindent
{\it Energy Generation and Energy Budgets}
\medskip

Other mechanisms can play a role as well in fueling the central heat
engine needed to balance the heat flow from the core to the cluster
halo.  Mass loss through stellar evolution (especially the much more
rapid stellar evolution of merger remnants) can indirectly heat a
cluster through the paradoxical effect of carrying off {\sl kinetic}
energy -- simply because the {\sl potential} energy carried off per
unit mass is much larger, and tilts the balance towards an effective
heating.  Similarly, the formation of a modest black hole can also
cause a heating of the cluster, through the selective eating of stars
on low-energy orbits near the hole (for both mechanisms, see the
review by Goodman 1992).

It is far from clear to what extent these various heating mechanisms
compete with each other in actual globular cluster cores.
Order-of-magnitude estimates indicate that they all can be
significant, depending on the precise conditions in the cores, as well
as on the nature of the stars.  For example, white dwarfs, neutron
stars and stellar-mass black holes are likely to produce energy by
dynamical binary formation and hardening, while main-sequence stars
and giants are likely to suffer physical collisions while attempting
to do so.

Whatever the detailed mix of energy sources in individual clusters may
turn out to be, binaries play a central role in the energy budget of a
globular cluster.  For example, observations of primordial binaries in
globular clusters indicate that the binary abundance in globular
clusters is not much smaller than that in the Galactic disk and halo
(as reviewed recently by Hut \etal\ 1992; see also Kaluzny \&
Krzeminski 1993 for additional binary detections in NGC 4372).  This
suggests that $\simgt 10$\% of the stellar objects in a cluster may be
binaries with an orbit of $\simlt 1$A.U., which implies a average
binding energy per binary of $\simgt 10$ times that of the average
kinetic energy of single cluster stars.

This simple reasoning leads to the astonishing conclusion that the
internal energy reservoir in binary binding energies may well exceed
the total amount of kinetic energy in the cluster as a whole (in the
form of center-of-mass motion of single stars and binaries).

The dominant role played by the internal degrees of freedom of
binaries in the overall energy budget already suggests that we'd
better provide an accurate treatment of binary star evolution, if we
want our overall cluster evolution to be believable.  This is the
topic of the next section.

\section{Ecological Reaction Networks}

With binary stars having locked up the bulk of the energy content of a
typical globular cluster, we cannot afford to neglect the
transformations in binary properties that take place in the course of
normal stellar evolution.  The reason is that stellar encounters do
not have a monopoly on changing the energy and angular momentum of
binaries; isolated binaries, too, have plenty of ways of changing
their appearance in complicated ways (for a fascinating account of an
ensemble simulation of these processes, see the contribution by Onno
Pols in these proceedings).

Even a partial list of some of the processes involved in isolated
binary evolution gives an idea of the complexity of the physics, such
as there are: tidal capture, magnetic breaking, gravitational
radiation, run-away mass transfer, and common envelope evolution.
Take into account the manifold perturbations and disruptions that can
occur when passing stars or binaries thicken the plot, and you see
what we are up against.  Clearly, the feed-back mechanisms between
stellar dynamics and stellar evolution in globular clusters play a
major role in the evolution of the cluster as a whole.  The term
`ecology', used by Douglas Heggie in his recent `news and views'
article in Nature (Heggie 1992), indeed captures the essence
of this interplay.

\bigskip
\noindent
{\it Blue Stragglers}
\medskip

In those clusters that have a relatively low central star density, as
well as in the outer areas of all clusters, blue stragglers can be
formed by mass overflow from an evolving star in a tight binary to the
(initially) less massive star (Pols, this volume).  In addition,
physical collisions between initially unrelated single stars must
produce blue stragglers as well, in the denser cluster cores, as was
first realized by Hills \& Day (1976).  Furthermore, encounters
between single stars and binaries are even more efficient in inducing
physical collisions between stars, as was pointed out by Hut \&
Verbunt (1983).

More detailed estimates by Krolik (1983), Krolik, Meiksin
\& Joss (1984), and Hut \& Inagaki (1985) confirmed the fact that
many thousands of stellar collisions must have taken place throughout
the history of our globular cluster system.  The feedback of these
merger remnants on the dynamical evolution of the cluster itself was
first taken into account by Lee \& Ostriker (1986) and Lee (1987).

Unfortunately, the present state of cluster modeling still does not
allow us to make significant improvements over the order-of-magnitude
estimates in the papers quoted above.  As discussed by Hut
\etal\ (1992), Fokker-Planck models have two intrinsic handicaps that make
them unsuitable for a quantitative modeling of the evolution of a blue
straggler population.  First, they are not set up to deal with the
separate evolution of internal and external degrees of freedom of the
binaries that play an important role in the formation and evolution of
blue stragglers.

The second problem stems from an introduction of a mass spectrum, as
well as a distinction between stars of different radii, such as
dwarfs, main-sequence stars, and giants.  The root of the problem here
is that a Fokker-Planck approach does not follow individual stars, but
rather distribution functions.  When the number of independent
parameters characterizing the distribution functions becomes too large,
there will be less than one star left in a typical cell in parameter
space --- something that clearly invalidates the statistical
hypothesis on which the Fokker-Planck approach is based.

The only solution seems to be to drop the statistical assumption, and
to revert to a star-by-star modeling of a globular cluster, through
direct $N$-body calculations.  Unfortunately, such calculations are
extremely expensive.

\section{The Hardware Barrier: Toward a Teraflops Speed}

Progress in increasing the number of particles in $N$-body simulations
of star clusters has been slow.  In the sixties, $N$ was measured in
the tens; in the seventies in the hundreds; and in the eighties in the
thousands.\footnote
{The 1993 record is a run with 10,000 particles, which reached core
collapse after spending a CRAY YMP month of CPU time spread out over
about a year (Spurzem \& Aarseth 1993).}
Peanuts indeed, compared to cosmological simulations with $10^6 \sim
10^7$ and more particles!  Why are these numbers so different?

The main reason is the fact that star clusters evolve on the time
scale of the two-body relaxation time at the half-mass radius,
$t_{rh}$.  In terms of the half-mass crossing time $t_{ch}$, $t_{rh}
\propto N t_{ch}$.  In addition, the use of tree codes or other codes
based on potential solvers is not practical, given the high accuracy
required.  This follows from the fact that an evolving star cluster is
nearly always close to thermal equilibrium.  Therefore, small errors
in the orbits of individual particles can give rise to relatively much
larger errors in the heat flux through the system.

As a result, the cost of a simulation of star cluster evolution scales
roughly $\propto N^3$, since $N$ crossing times are needed for each
relaxation time, and all $N^2$ gravitational interactions between the
stars have to be evaluated many times per crossing time.  In contrast,
most simulations in galactic dynamics span a fixed number of crossing
times, and can employ an algorithm with a computational cost scaling
$\propto N\log N$.  It is this difference, of roughly a factor $N^2$,
which has kept cluster modeling stuck in such a modest range of
$N$-values.

\bigskip
\noindent
{\it What it Takes}
\medskip

The above rough estimates, although suggestive, are not very reliable.
In an evolving star cluster, large density gradients will be formed,
and sophisticated integration schemes can employ individual time steps
and various regularization mechanisms to make the codes vastly more
efficient (see the next section).  One might have hoped that the above
scaling estimates might have turned out to be too pessimistic.
Indeed, empirical evidence based on runs in the range $100 < N < 1000$
seemed to indicated a scaling of the computational cost $\propto
N^{1.6}$, rather than $N^2$, per crossing time (Aarseth 1985).

However, a detailed analysis of the scaling behavior of various
integration schemes (Makino \& Hut 1988) showed that the relatively
more mild empirical scaling must have been an artifact of the still
rather small $N$ values it was based on.  Instead, the best asymptotic
scaling reported by Makino \& Hut, based on a lengthy analysis of
two-timescale methods, resulted in a computational cost $\propto
N^{25/12}$ per crossing time in a homogeneous system, as well as in an
isothermal density distribution.  For steeper density distributions,
with the velocity dispersion increasing towards the center, they found
even slightly higher cost estimates.

With this staggering scaling of the computational cost, slightly worse
than $\propto N^3$, it would seem worthwhile to look for some type of
approximate method, in which part of the star cluster is modeled in a
statistical fashion.  Indeed, such a hybrid approach, using a
combination of Fokker-Planck methods and direct $N$-body integration,
was carried out by McMillan and Lightman (1984ab) and McMillan (1985,
1986).  This approach proved to be successful in modeling a star
cluster around the moment of core-collapse.  However, these
simulations could not be extended significantly beyond core collapse,
for lack of computer time.

When Hut, Makino \& McMillan (1988) applied the analysis by Makino \&
Hut to McMillan's hybrid code, as well as to a variety of other
algorithms, they reached a pessimistic conclusion.  In order to model
even a modest globular cluster with $10^5$ stars, even with the
theoretically most efficient integration scheme, would carry a cost of
several Teraflops-days (a Teraflops equals $10^{12}$ floating point
operations per second, and a Teraflops-day therefore corresponds to
$10^{17}$ floating point operations).  More realistically, they
concluded, for a relatively simple and vectorizable or parallelizable
algorithm, the computational cost would lie in the range of
Teraflops-months.

\bigskip
\noindent
{\it How to Get There}
\medskip

Available computer speed increases by a factor of nearly two each
year, or a factor of $300\sim1000$ per decade.  Ten years ago, the
speed of personal computers (or a per-person-averaged speed of a VAX),
was measured in kflops, while current workstations are measured in
terms of Mflops.  Extrapolating, by the year 2013 we can expect each
of us to have a multi-Teraflop machine on our desk, enabling us to
core-collapse a globular cluster in a week or so.

If we look at the speed of the fastest supercomputers available, a
similar scaling holds.  Ten years ago, astrophysicists could have
occasional access to a supercomputer, but when averaged to a sustained
speed, it would boil down to only a fraction of a Mflops speed.  And
indeed, by now the best one can hope to obtain from a NSF
supercomputer center is an average speed of a fraction of a Gflops.
It would seem that the evolution of even a modest globular cluster
would not be possible until some time around the year 2003.

Fortunately, we do not have to wait that long.  Following the rather
pessimistic conclusions of Hut \etal\ (1988), a project was started to
construct special-purpose hardware for $N$-body calculations, by a
small group of astrophysicists at Tokyo University (Sugimoto \etal\
1990).  This GRAPE project (from `GRAvity PipE') centers on the
development of parallel Newtonian-force-accelerators, in analogy to
the idea of using floating-point accelerators to speed up
workstations.  During the last three years, a variety of GRAPE
versions has been completed (and some of them distributed to other
locations outside Japan), and used successfully for several
astrophysical calculations (see, e.g., Funato, Makino \& Ebisuzaki
1992).

The next major step in this project will be the development of a
Teraflops speed high-accuracy special-purpose computer (the HARP, for
`Hermite AcceleratoR Pipeline'; Makino, Kokubo \& Taiji 1993), in the
form of a set of a few thousand specially designed chips, each with a
speed of several hundred Mflops.  This machine is expected to be
available some time next year.

\section{The Software Barrier: Recursive Transformations}

The shortest length and time scales of interest in globular cluster
evolution are posed by the closest encounters of the most compact
types of stars.  This leads us to a near-grazing encounter of two
neutron stars, an event with a duration of order of a few
milliseconds.  Compared with the age of a globular cluster, $10^{10}$
years, we have a discrepancy in time scales of no less than 20 orders
of magnitude!

A similar problem shows up when we compare length scales.  The
diameter of a neutron star, expressed in units of the tidal radius of
a globular cluster, is of order $10^{-15}$.  At first sight, modeling
a globular cluster on a star-by-star basis would seem pretty hopeless.
The fact that we are able to contemplate such an enterprise at all is
largely due to the ingenuity and persistence of Sverre Aarseth, who
has provided a framework in which to tackle these seemingly
unsurmountable problems (together with efficient implementations).
The key words for facing up to the length scale and time scale
problems are individual timesteps and local coordinate
transformations.

\bigskip
\noindent
{\it Time Scale Problem}
\medskip

In the course of the sixties, it was gradually realized how important
binaries are in determining the dynamics of star systems large or
small -- and at the same time how devastating their presence was for
one's computer budget.  Even in a modest 10-body simulation, the
formation of a single hard binary is guaranteed to slow down the speed
of the whole simulation by orders of magnitude, when using one of the
standard integration schemes for coupled differential equations.

The reason is that even with variable time steps, these standard
schemes force all stars to share the same time step size.  As a
result, a typical timestep size of a few hundred years in a loose
association of ten stars can be reduced to a month or less with the
first moderately tight binary appearing on the scene.

The answer to this problem, provided by Aarseth (\cf Aarseth 1985),
was the introduction of individual time steps.  Instead of viewing a
star cluster with three-dimensional eyes, as made up of a set of mass
points in space, Aarseth took a four-dimensional view, in which each
star was replaced by an orbit segment.  Whenever the time had come for
a particular star to move, it could determine the gravitational
acceleration by all its neighbors by asking them to slide along their
orbit segments (implemented as polynomial approximations) to the
desired point in time, even though each of them in turn had computed
its own positions, velocities, and accelerations only at earlier times.

This single algorithmic improvement did more to speed up star cluster
calculations than decades worth of hardware speed improvement.  Even
on today's fastest machines, it would be difficult to reproduce
Aarseth's calculations of twenty years ago without using individual
timesteps.

\bigskip
\noindent
{\it Length Scale Problem}
\medskip

The large range in length scales, although somewhat smaller than the
range in time scales, has turned out to pose a more significant
problem for simulations of star cluster evolution.  In contrast to the
time scale problem, which simply slows things down to a crawl, the
length scale problem can easily make a whole calculation meaningless.

The main problem lies in round-off errors.  When we take a binary
consisting of two neutron stars, moving in the outskirts of a globular
cluster, there is no need to follow its internal orbit as long as
other perturbers are far away (another significant software
improvement provided by Aarseth).  However, in the rare case that a
third star would interact with that type of binary, we would have to
follow all three point-particles numerically, during the time of the
interaction.

And here the problem shows up most clearly: for the position vectors
of the two neutron stars, with respect to the center of mass of the
cluster, the first fifteen digits could turn out to be the same, as we
saw above.  Alas, even a double precision (64-bit) representation of
floating point numbers typically carries a mantissa of only fifteen
significant digits.  So much for integrating equations of motion.

Admittedly, this is an extreme case, but it makes a point: loss of
accuracy due to round-off when subtracting large numbers is a very
serious worry in $N$-body calculations.  Aarseth's answer was to
implement a series of ever more intricate treatments of two-body,
three-body and more complex multiple encounters.  These treatments are
largely based on Kustaanheimo-Stiefel regularization procedures, in
which the three-dimensional Kepler singularity is `unfolded' by a
coordinate transformation to four dimensions, mapping the
three-dimensional Kepler problem into that of a four-dimensional
harmonic oscillator through the inverse of one of the Hopf maps from
$S^3$ to $S^2$ (\cf Stiefel \& Scheifele 1971), introducing a U(1)
gauge symmetry in the process.

\bigskip
\noindent
{\it Code Development}
\medskip

Aarseth has implemented various other algorithmic improvements as
well, most notably the Ahmad-Cohen neighbor scheme, in which the
individual time step scheme is further refined to a two-time-scale
approach (by a more frequent calculation of nearby interactions,
compared to remote interactions).  The resulting code, NBODY5, that
includes all these improvements, has been the tool of choice for any
type of detailed star cluster $N$-body simulation, for well over a
decade.  The community owes a debt of gratitude to the generosity and
dedication of Sverre Aarseth, who not only has made his codes widely
available, but has made himself available as well -- for friendly
advice as much as for code repair and maintenance.

All these developments notwithstanding, there are still some major
problems facing us, if we want to carry out a star-by-star $N$-body
simulation of globular cluster evolution.  First of all, if the past
can offer any guidance, increasing the number of particles by a
significant factor is guaranteed to uncover new algorithmic
bottlenecks, requiring new solutions and fine-tuning.  The present
state of Aarseth's codes is the outcome of an evolutionary process of
successive attempts to deal with increasingly harder problems, posed
by the increase in complexity of the systems to which it has been
applied.  This process is likely to continue for quite a while.

Secondly, the large set of heuristic improvements, valuable as they
have been in making calculations possible in the first place, pose a
formidable problem to the implementation of stellar evolution recipes.
Even simple questions such as the value of the distances between
various particles at a given time becomes less straightforward as it
may seem at first, when one realizes that one has to trace these
distances in the guise of their four-dimensional transformations in
which time itself is an extra coordinate, given as a function of the
independent integration parameter.  Add to this the veritable
complexities in the Ahmad-Cohen bookkeeping of the two time scales
used (separately for each individual particle, introducing a
multiplicity of orbit segment representations), and the outline
of the daunting task one is facing begins to appear.

Thirdly, and equally important, it would seem less than ideal if all
simulations in an entire field of astrophysics would be continued with
a single code, without any independent form of comparison or
calibration.  In a sense, winning the star cluster $N$-body space
race hands-down, twenty years ago, has been a mixed blessing for
Sverre Aarseth.  He has gained a lot of friends, but at the same time
has lacked any significant form of competition.  It would seem high
time for someone to explore alternative approaches to $N$-body code
writing.

\bigskip
\noindent
{\it A Recursive Approach}
\medskip

For all these reasons, it has been clear for many years that an
independent approach was called for.  With the prospect of teraflops
speeds becoming available in the near future, Jun Makino, Steve
McMillan and I decided that the time had come to take on the somewhat
daunting task of engaging Sverre Aarseth in an amicable competition.
Apart from the adviced to `stay tuned', I briefly sketch what lies at
the core of our approach (Hut \etal\ 1993).

The central idea we have introduced is that of recursive coordinate
transformations.  The underlying notion is that of hierarchical
simplicity.  Rather than giving a special treatment to each of a
number of different closely interacting groups (binaries, triples,
binary-binary encounters, etc.), we use a recursive approach to split
up an interacting group of stars in subgroups, down to the level of
individual stars.  The advantage is twofold: we can now handle
arbitrary $k$-body subsystems, with $k>4$ as well; and we can treat
especially tight subgroups that may appear deep inside an already
tight group (or even tighter sub-subgroups).

Most stars are unaffected by these special treatments, and are
simply represented as leaves of a flat top-level tree.  The stars
that take part in close encounters or are members of multiple star
systems, however, play a crucial role in cluster simulations.  The
bulk of NBODY5, for example, is concerned with special treatments of
closely interacting groups of $\leq 4$ stars.  In addition, in most
simulations so far most of the computer time has been taken up by the
integration of the equations of motion of this stellar minority.

In our treatment of such groups of closely interacting stars, the
center of mass of the group is represented by a node in the top-level
tree.  This node in turn serves as the local root node for a binary
tree (a tree with two branches per node), in which each star forms a
leaf.  The structure of this tree is changed dynamically, to guarantee
that the tree structure closely reflects, at any time, the
configuration of the (sub)groups of the stars.  Our approach somewhat
resembles that taken by Jernigan and Porter (1989), the major
differences being that we limit ourselves to small subsets of stars,
and at least for the time being forgo any fancy regularization
technique.

Whether we can get away with a bare-bones set of recursive coordinate
transformations, rather than four-dimensional regularizations, remains
to be seen.  But in any case, our recursive approach to a dynamical
maintenance of a tree configuration is likely to alleviate our task of
providing a clean interface between the stellar dynamics and the
stellar evolution parts of our code.

Clearly, the art of $N$-body modeling is a vast subject that still
leaves room for many novel approaches.  Extending its realm of
application to the dynamics of $10^5$ stars with, say, $10^4$
primordial binaries, is a sure-fire way to stimulate further
exploration.

\section{The Physics Barrier: Recipes for Stellar Evolution}

What will happen when a W Uma contact binary encounters another double
star, say a red giant in orbit around a black hole (`Bambi meets
Godzilla'), after the four stars begin a chaotic four-body dance?  It
is not inconceivable, in the dense environment of a post-collapse
cluster core, that an unrelated triple system will saunter by, say in
the form of a white dwarf closely orbiting a neutron star, having just
captured a main-sequence star in a wide elliptic orbit.

Farfetched, such a scenario, of a tightly-interacting seven-body
system?  Not really.  With $\simgt 10^3$ stars in a globular cluster
core, during a post-collapse period of $\simgt 10^9$ years, and a core
crossing time of $\simlt 10^4$ years, occasional traffic jams like the
one sketched above are bound to happen.

A glance at the number of free parameters involved in such a complex
encounter will dispel any thought of an approach based on some type of
table lookup.  Tabulating (or giving functional fits to) equal-mass
three-body encounters in the point-mass limit is certainly doable
(Heggie \& Hut 1993), but in the much more complex cases such as the
one sketched above such an approach simply won't work.  To avoid our
computer code giving up in despair, we are aiming at constructing a
hierarchical set of recipes, which hopefully will handle encounters of
the type sketched above.

But, one cannot help wondering, do we really need to model a globular
cluster on this level of detail?  The answer is simple: we do.  The
reason is that there is no simpler simulation worth doing, because of
the large gap between an equal-mass point-particle simulation of
cluster evolution (unrealistic, but at least consistent), and a more
realistic simulation.  This is an important point, not generally
appreciated.  Let me be more specific.

\bigskip
\noindent
{\it Consistency}
\medskip

As soon as we introduce a mass spectrum in a star cluster simulation,
we will see that the heavier stars start sinking toward the center, on
the dynamical friction time scale, shorter than the two-body
relaxation time by a factor proportional to the mass of individual
heavy stars.  The reason is that relaxation tends toward equipartition
of energy, which implies that heavier stars will move more slowly and
therefore gather at the bottom of the cluster potential well.

If stars would live forever, there would be a large overconcentration
of heavy stars in the core of a star cluster.  However, in reality
there is a important counter-effect: heavy stars burn up much faster
than lighter ones.  They may or may not leave degenerate remnants, that
may or may not be heavier than the average stellar mass in the cluster
(a quantity that also decreases in time).  Clearly, it would be
grossly unrealistic to introduce a mass spectrum without removing most
of the mass of the heaviest stars on the time scale of their evolution
off the main sequence and past the giant branches.

Another reason for introducing finite life times for stars comes from
abandoning the very restrictive point mass model for stars.  As soon
as we do that, giving our stars a finite radius will give rise to
stellar collisions.  The heavier stars produced in the collision of two
turn-off stars will burn up in one or two billion years.  Again, we
have to take this into account to be consistent, especially since the
merger products themselves are prime candidates for further merging
collisions.

The need to let many stars shed most of their mass, together with the
fact that most of the energy in a globular cluster is locked up in
binaries, poses a formidable consistency problem.  Since binaries play
a central role in cluster dynamics, consistency requires that we
follow their complex stellar evolution, which involves mass overflow
(which can be stable or unstable, and take place on dynamical or
thermal or nuclear time scales) and the possibility of a phase of
common-envelope evolution.  On top of all that, we will have to find
simple recipes for the hydrodynamic effects occurring in three-body and
four-body reactions, and in occasional $N>4$ reactions, as indicated
above.

To sum up: there does not seem to be a half-way stopping point, at
which we can expect to carry out consistent cluster evolution
simulations.  Either we study the interesting but unrealistic
mathematical-physics problem of an equal-mass point particle model, or
we opt for the realistic model with some set of stellar-evolution
bells and whistles.  The only question is: what is the simplest set
that is still consistent?

\bigskip
\noindent
{\it Getting Started}
\medskip

Another way of posing the question is: how to mimic some form of
stellar evolution that is utterly simple but not totally silly.  We
would be more than happy with a simple toy-model for starters,
something which makes errors of factors-of-a-few in many places,
without being altogether ridiculous (no order-of-magnitude errors).
{}From there on, we can then make further progress through a series of
closely placed stepping stones, by further improving each of the many
ingredients in the recipes hinted at above.

However, to build a not-altogether-silly toy model is far from
trivial.  The main problem is that a simulation with 100,000 stars
will give rise to so many different types of interactions that human
intervention will become impractical.  The code will have to contain
some rudimentary knowledge about each foreseeable (and probably as yet
unforeseen!) encounter.

Once a minimum treatment of stellar evolution effects is included in
future simulations, we have to face the question of the extent to
which the initial rudimentary modeling can be improved.  Here the
prospects may well be limited, due to fundamental uncertainties
posed by such processes as common-envelope evolution.  Until detailed
three-dimensional hydrodynamical modeling (including a full radiative
treatment!) becomes available for such cases, there seems to be little
reason to extend our treatment much beyond the simplest type of
consistent implementation.

The scope for stellar evolution modeling thus seems well-determined by
limitations to the allowed complexity on either end of the scale.  I
am confident that in a couple years time we will succeed in an
implementation of this rather well-determined set of recipes.  By
then, the full wealth of observations of X-ray binaries, millisecond
pulsars and blue stragglers can be brought in, and compared with our
simulations.  This will finally allow us to obtain a coherent picture
of the so-far-elusive quantitative aspects of the structure and
evolution of globular clusters.

\section{Kilobyte Needles in Terabyte Haystacks}

Generating data is only half the job in any simulation.  The other
half of the work of a computational theorist parallels that of an
observer, and lies in the job of data reduction.  As in the
observational case, here too a good set of tools is essential.  And
not only that: unless the tools can be used in a flexible and coherent
software environment, their usefulness will still be limited.

Three requirements are central in handling the data flow from a
full-scale globular cluster simulation: modularity, flexibility, and
compatibility.  We have started to put together a software
environment, Starlab (Hut \etal\ 1993), that incorporates these three
requirements.  To some extent, Starlab is modeled on NEMO, a stellar
dynamics software environment developed six years ago at the Institute
for Advanced Study, for a large part by Josh Barnes with input from
Peter Teuben and me, and has subsequently been maintained and extended
by Peter Teuben.

Starlab is different from NEMO mainly in the following areas: it
emphasizes the use of UNIX pipes, rather than temporary files; its use
of tree structures rather than arrays to represent $N$-body systems; and
its guarantee of data conservation -- data which are not understood by
a given module are simply passed on rather than filtered out.

\bigskip
\noindent
{\it Modularity: A Toolbox Approach}
\medskip

We have followed the UNIX model of combining a large number of small and
relatively simple tools through pipes.  This allows a quick and
compact way of running small test simulations.  For example, a study
of relaxation effects in a cold collapse could be done as follows:

\def\codes{
\parindent=0pt
\obeylines
\tt
\medskip
}

{\codes
mkplummer -n 100 | freeze | leapfrog -t 2 -d 0.02 -e 0.05 | lagrad
}
\medskip

Here {\tt \ mkplummer\ } creates initial conditions for a 100-body system,
according to a Plummer model distribution.  The resulting data are
piped into the next module, {\tt \ freeze\ }, which simply sets all
velocities to zero, while preserving the positions.  Following that,
the data are read in by the leapfrog integrator, which is asked to
evolve the system for a period of 2 time units, with a stepsize of
0.02 time units, and a softening length of 0.05 length units.  Finally,
the resulting data are piped into a module that makes a plot of the
Lagrangian radii of various percentiles of the system.

\bigskip
\noindent
{\it Flexibility: Structured Data Representation}
\medskip

Each snapshot of a $N$-body simulation can be stored in a file in a
standard format, with a header indicating the nature of the snapshot.
In addition, a list of all the commands used to create the data is
stored at the top of the file, together with the time at which the
commands were issued, so as to minimize the uncertainty about the
exact procedures used.  Each individual body is presented as a node in
a tree, constructed so as to reflect the presence of closely
interacting subsystems and their internal structure.

Each body has several unstructured `scratch pads', in which each
application program can write diagnostics or other comments describing
particular occurrences during the integration.  This has proved to be
extremely useful, by allowing various forms of data reduction to take
place already during the run.  Especially during complicated
interactions involving stellar dynamics, stellar hydrodynamics, and
stellar evolution effects, a free-format reporting system, tied to the
individual interacting objects, will be very helpful in allowing a
reconstruction of episodes of greatest interest.

\bigskip
\noindent
{\it Compatibility: Unfiltered Piping}
\medskip

The internal data representation of each module is such that
unrecognized quantities or comments are stored internally, in the form
of character strings.  They are reproduced at the time of output, at
the correct position, preserving their correspondence with the initial
bodies they were associated with (some of which may have collided and
merged).  This allows the use of an arbitrary combination of pipes with
the guarantee that no data or comments will be lost.

For example, in the commands

{\codes
evolve | mark\_core | HR\_plot | evolve
}
\medskip

the first module evolves the system, integrating the equations of
motion, while also following the way the individual stars age and
interact hydrodynamically.  The second program computes the location
and size of the core of the star cluster, and marks those particles
that are within one core radius from the center.  The third module
plots a Hertzsprung-Russell diagram of the star cluster (perhaps using
special symbols for the core stars), before passing on the data once
more to the module that evolves the whole system.  For this to work,
the {\tt \ mark\_core\ } program needs to preserve the stellar
evolution information, even though it only `knows' about the stellar
dynamical part of the data.  Similarly, the {\tt \ HR\_plot\ } program
needs to preserve the dynamical data.

\section{Summary and Outlook}

The study of star cluster evolution has seen steady progress
throughout the last several decades.  Analytic estimates dating back
from the late thirties, together with numerical simulations in the
late sixties and seventies, have provided the stage for the
break-through in our understanding in the early eighties, when the
evolution through core collapse and beyond began to be understood.

In the ten years following this break-through, the various modeling
techniques have been pushed to the limit of their applicability.  We
are now facing three barriers separating us from our goal of an
adequate treatment of star cluster evolution: a star-by-star $N$-body
modeling, including stellar evolution as well as stellar dynamics.

Two of the three barriers are related to the stellar dynamics part of
the problem: getting the raw speed to do the simulations, and
developing the algorithms necessary to utilize that speed.  The third
barrier is related to the implementation of an utterly simple (but not
too simple) treatment of the various stellar evolution effects
mingling with stellar dynamics.

The identification and initial exploration of these barriers begun in
earnest five years ago, with a detailed analysis of the requirements
of a star-by-star simulation (Makino \& Hut 1988; Hut, Makino \&
McMillan 1988).  The resulting specification of the necessity of a
computer speed of order a Teraflops at the time made our goal seem
remote.  And indeed, it is unreasonable to expect full-time access to
a Teraflops computer before some time early in the next century.

Fortunately, we do not have to wait a decade.  Thanks to the hardware
development of the Tokyo group (Makino \etal\ 1993), we now have a
definite time table for our first star-by-star globular cluster
simulations, with a number of stars in the range of $50,000$ --
$100,000$: this is expected to take place some time in 1994.

\section{Acknowledgments}

I thank Sverre Aarseth, Jun Makino, and Steve McMillan for comments on
the manuscript.

\end{document}